# Oxygen Evolution Reaction on Perovskites: A Multieffect Descriptor Study Combining Experimental and Theoretical Methods

Xi Cheng, Emiliana Fabbri, Yuya Yamashita, Ivano E. Castelli, Baejung Kim, Makoto Uchida, Raphael Haumont, Inés Puente-Orench, Thomas J. Schmidt

Abstract

The correlation between ex situ electronic conductivity, oxygen vacancy content, flat-band potential ($E_{fb}$), and the oxygen evolution reaction (OER) activity for a wide range of perovskite compositions [$La_{1-x}Sr_xCoO_{3-\delta}$ series (with $x$ = 0, 0.2, 0.4, 0.6, 0.8), $LaMO_{3-\delta}$ series (M = Cr, Mn, Fe, Co, Ni), $Ba_{0.5}Sr_{0.5}Co_{0.8}Fe_{0.2}O_{3-\delta}$ (BSCF), and $PrBaCo_2O_{6-\delta}$ (PBCO)] are investigated experimentally and theoretically. It is found that all of these parameters can affect the OER activity; however, none of them alone play a crucial role in determining the electrocatalytic activity. The correlation of one single physicochemical property with the OER activity always presents deviation points, indicating that a limitation does exist for such 2-dimensional correlations. Nevertheless, these deviations can be explained considering other physicochemical properties and their correlation with the OER activity. Hence, this work aims in simultaneously linking the OER activity with several physicochemical materials properties. The concept of the OER/multidescriptor relationship represents a significant advancement in the search and design of highly active oxygen evolution catalysts, in the quest for efficient anodes in water electrolyzers.

# Introduction

In the course of replacing fossil-fuel-based energy technologies the development of energy storage systems is crucial to mediate the variable nature of energy generation from renewable resources. Within this scenario, water electrolysis technologies have been the center of the spotlight. Particularly the reaction at the anode side of water electrolyzers, the oxygen evolution reaction (OER), has achieved a lot of research and development attention since it generally requires high overpotentials that limit the energy efficiency of the process.[1,2] However, despite the intense research efforts dedicated to the understanding of the OER mechanism on the surface of oxide materials, there is still little comprehension of the individual oxygen evolution reaction steps, and to date, there is not a unanimous consensus on the reaction mechanism.[1] The study of the OER mechanism is particularly complex because other electrochemical reactions can take place on the surface of oxide catalysts concurrently to the OER, such as the metal oxide dissolution or the lattice oxygen evolution reaction (LOER).[3−8] Therefore, the understanding of the OER mechanism would require disentangling it from the metal dissolution and the LOER mechanisms, which is unquestionably a difficult task. For this reason, the most common approach is to assume that the same reaction mechanism takes place for all the oxide catalysts and identify descriptors, which can be defined as the catalyst physicochemical properties affecting the measured OER activity. Therefore, activity descriptors do not point directly to a specific OER mechanism but rather help in proficiently and effectively predicting the most OER-active oxide catalysts. However, the understanding of the true activity descriptor for a class of catalysts might also help in elaborating a more realistic reaction mechanism than that initially assumed. For example, the study of Halck et al.[9] shows that by doping $RuO_2$ with Ni or Co deviation from the volcano plot correlating the OER overpotential and the difference between the surface binding energies of O* and HO* reaction intermediates ($\Delta G_{O^*} - \Delta G_{HO^*}$) could be observed. This deviation has been explained by a different reaction mechanism involving the presence of a cocatalyst, which activates a proton donor–acceptor functionality on the conventionally inactive bridge surface sites.[9] However, the study of the electrochemical stability of $IrO_2$ doped with Ni catalysts shows a correlation between the Ni dissolution and the OER activity.[10] The leaching of the transition metal from the mixed oxides leads to a more active $IrO_2$ surface area for the OER rich in hydroxyl groups which have been suggested as the activity descriptor for this class of material, indicating that the understanding of the OER mechanism is deeply linked to the investigation of the catalyst properties.

Perovskite oxides ($ABO_{3-\delta}$) with alkaline or rare-earth cations in the A-site and first-row transition-metal cations in the B-site are attractive candidates as electrocatalysts for the oxygen evolution reaction (OER) in alkaline electrolytes.[1] Their physicochemical properties as well as their catalytic activity could be significantly influenced by partial substitution of the A- and/or B-site by other elements resulting in $(A_xA'_{1-x})(B_yB'_{1-y})O_{3-\delta}$ compositions. Moreover, it has been reported that Co-based double perovskites with a formula unit of $LnBaCo_2O_{6-\delta}$ (Ln = Pr, Sm, Gd, and Ho) also present an interesting OER performance.[2] The attractive possibility given by the perovskite family to investigate a wide range of oxide compositions also asks for design principles able to proficiently and effectively predict the most OER-active perovskite catalysts. In this context, extensive studies have been performed to understand the key material's physicochemical properties influencing the perovskite OER activity,[11-19] with the further purpose of improving known materials or predicting new active materials. At present different activity descriptors for the perovskite OER activity have been proposed such as the number of d-electrons,[11,17] the $e_g$ band filling of the transition-metal cations,[14] the difference between the surface binding energies of O* and HO* reaction intermediates ($\Delta G_O^* - \Delta G_{HO}^*$),[12] the oxide formation energy,[16] and the accumulation of the magnetic moment on the conduction plane atoms.[19] However, there is no consensus on the bulk physicochemical property that univocally predicts fast OER kinetics for a wide range of perovskite compositions. Indeed, even though good correlations between the OER activity and a material physicochemical property can be obtained for a certain perovskite series, when a wide range of perovskite compositions is considered most of the proposed catalytic activity descriptors would fail for some specific materials. This suggests that a single material physicochemical property might not be sufficient to describe a complex electrochemical scenario where at least three different reaction pathways, i.e., OER, oxide dissolution, and LOER, are simultaneously occurring. However, if a single descriptor might fail in predicting the key property for achieving a high OER activity, a combination of several descriptors could lead to a more comprehensive understanding of the crucial parameters influencing the OER activity of this complex class of materials.

Thus, in the present work, we have refined the search for an OER activity descriptor not only considering a single physicochemical material property but also trying to correlate the OER activity of a wide range of perovskite compositions to a network of multiphysicochemical properties. In particular, we have investigated the La$_{1-x}$Sr$_x$CoO$_{3-\delta}$ (with $x$ = 0, 0.2, 0.4, 0.6, 0.8) and LaMO$_{3-\delta}$ (M = Cr, Mn, Fe, Co, Ni) series, and two complex perovskites: Ba$_{0.5}$Sr$_{0.5}$Co$_{0.8}$Fe$_{0.2}$O$_{3-\delta}$ (BSCF) and PrBaCo$_2$O$_{6-\delta}$ (PBCO).

In our previous study,[20] we have demonstrated that, by varying the A-site composition in the La$_{1-x}$Sr$_x$CoO$_{3-\delta}$ series (with $x$ = 0, 0.2, 0.4, 0.6, 0.8, 1), we could tune the oxide bulk structure, electronic properties, and electronic conductivity. With the help of density functional theory (DFT) calculations, we could link these physicochemical properties, and we found that a linear OER activity/conductivity correlation for this perovskite series exists. However, in the present study a nonlinear OER activity/conductivity has been observed by extending the same OER correlation to a wide range of perovskite compositions, demonstrating that in addition to conductivity other perovskite properties play an important role in the OER activity. Therefore, in addition to the bulk/electronic structure and the conductivity, we have considered two additional physicochemical properties: the oxygen vacancy content and the flat-band potential ($E_{fb}$). Some studies have recently reported that the presence of oxygen vacancies in the lattice could strongly affect the electrocatalytic activity of perovskites[3] and of other types of metal oxide catalysts.[21,22] Furthermore, we have recently proposed that highly active perovskite catalysts, such as BSCF, develop a self-assembled and highly active superficial oxyhydroxide layer rich in the perovskite B-site cations because of the simultaneous occurrence of OER, LOER, and cation dissolution.[3,4] It was further suggested that perovskite structures with a high content of oxygen vacancies might be more prone to undergo LOER and thus develop a highly active oxyhydroxide surface layer.[3] Therefore, even though the OER mechanism on perovskite catalysts is actually directly linked to the formation of a superficial oxyhydroxide layer, specific bulk properties, such as the oxygen vacancy content, are still essential for the formation of the self-assembled oxyhydroxide layer. However, a thorough discussion about the effect of oxygen vacancies on the OER activity for a wide range of perovskite compositions is still missing. In particular, the study of the oxygen vacancy content in OER perovskite catalysts by using the neutron diffraction method has been rarely reported, even though neutron diffraction is a very accurate method to determine the oxygen vacancy concentration in oxides. Thus, in this work, by the use of neutron diffraction, an OER/oxygen vacancy content correlation has been investigated for two different perovskite series (La$_{1-x}$Sr$_x$CoO$_{3-\delta}$ and LaMO$_{3-\delta}$ series) and two complex perovskites (BSCF and PBCO). The obtained results show that high oxygen vacancy content favors OER activity. Furthermore, the OER/oxygen vacancy trend could explain some of the deviation points in the OER/conductivity correlation; however, the OER/oxygen vacancy trend itself presents some deviations without an explanation. This orients us to look for a further physicochemical property able to fill the correlation gaps, particularly the flat-band potential ($E_{fb}$). The flat-band potential can be determined through the Mott–Schottky relation by calculating the electrode capacitance from the data obtained by impedance spectroscopy (IS) measurements at different applied potentials.(23) For a p-type semiconductor, which is the case for all the perovskites studied in the present work, when the applied potential is higher than the flat-band potential, an electron hole accumulation layer is formed on the catalyst surface region. It has been expected that the accumulation layer favors the charge transfer, and consequently triggers the OER.(23) Indeed, a correlation between the OER activity and the perovskite flat-band potential does exist, supporting our initial hypothesis. Moreover, the OER/flat-band potential correlation allows us to explain the deviations observed in the two other correlations mentioned above.

Concerning BSCF perovskite, it has been reported that BSCF as a single material electrode displays modest OER activity, while by developing composite electrodes made of BSCF and carbon, an improved catalytic activity and selectivity can be achieved.(24−26) Our previous study on BSCF/carbon composite electrodes by X-ray absorption near-edge spectroscopy (XANES) demonstrated that the presence of carbon induces a reduction of the Co oxidation state during the composite electrode preparation process, inducing an increase in the OER activity.(24,25) However, how the Co reduction affects the OER activity is still not clear. Hence, here we take advantage of the OER/descriptor study to find how the reduction of the Co cation oxidation state modifies the physicochemical properties of BSCF and how these properties affect its OER activity.

Finally, the present study clearly shows that even though deviation points are present if the OER activity is correlated to one single perovskite catalyst property, these deviations can be explained considering other material physicochemical properties and their correlation with the OER activity. Indeed, the proposed activity descriptors are not isolated from each other, but they are mutually correlated to the OER activity allowing the building up of an OER/multidescriptor correlation. Therefore, the present work is not just a simple assembly of different descriptors but rather an OER/multidescriptor study that demonstrates the limitation of a 2D correlation (i.e., relation of the OER activity to one single physicochemical property) and proposes the concept of a multidescriptor correlation; i.e., it is not a single material property that correlates with its OER activity, but only a set of material properties allow interpretation of its OER activity.

## Experimental Section

Material Preparation

Perovskite powders were synthesized with a modified sol–gel process. In brief, stoichiometric quantities of oxides or nitrate salt of A-site cation precursor [$La_2O_3$ (Aldrich, 99.9%), $Sr(NO_3)_2$ (Aldrich, 99%), $Ba(NO_3)_2$ (Aldrich, 99%), and $Pr_6O_{11}$ (Aldrich, 99.9%)] and nitrate salt of B-site cation precursors [Cr, Mn, Fe, Co, Ni (Aldrich)] were dissolved in an aqueous solution of 0.2 M nitric acid. Citric acid was used as a chelating agent in a 2:1 ratio with respect to the total metal cations (for $LaNiO_{3-\delta}$, citric acid and EDTA were used in a 2:1:1 ration with respect to the total metal cations). After the acquisition of a transparent solution, the pH was adjusted between 9 and 10 by $NH_4OH$ additions. The solution was then heated under stirring to evaporate water until it changed into a viscous gel and finally ignited to flame, resulting in ash. For the acquisition of a single phase oxide material, the perovskite powder was calcined in air (800–1050 °C for 2–10 h according to different oxide). Calcined BSCF was further functionalized by mixing BSCF and functionalized acetylene black carbon ($AB_f$) in a 5:1 ratio in isopropanol and then ultrasonicating the suspension for 30 min.(25) The $BSCF/AB_f$ suspension was centrifuged at 1000 rpm for 1 min, which allows precipitation of the BSCF component but not the $AB_f$. Successively the suspended $AB_f$ in isopropanol was removed, and finally, fresh isopropanol was added. The same procedure was repeated several times until no carbon in suspension was observed. The $AB_f$-free $BSCF/AB_f$ powder was then dried in air at 60 °C and denoted as $(BSCF/AB_f)_{centrifuged}$.

Structural and Chemical Characterization

Phase identification for perovskite powders was carried out by X-ray diffraction (XRD, Bruker D8 system) with Cu Kα polychromatic radiation (λ = 0.154 18 nm) in a Bragg–Brentano geometry. Neutron powder diffraction patterns were collected at 300 K on a D1B goniometer on a thermal source (λ = 0.128 nm) using the Institut Laue-Langevin reactor facility (ILL, Grenoble, France). Rietveld structural refinements have been achieved on the whole diffraction pattern, using Fullprof software.(27) The principle of the Rietveld method is to minimize the difference between a calculated profile and the observed data. Calculated intensity is adjusted with several variables, especially the positions of each atoms (later used for determinate lengths and angles of chemical bonds) and the average rate of oxygen in the $ABO_{3-\delta}$ oxide formula (later used for determination of the value of oxygen vacancies δ and therefore the effective charge of the B cation).

The specific surface area of the powder was determined by Brunauer–Emmett–Teller (BET) analysis. For X-ray absorption spectroscopy (XAS) measurements, catalyst powders were pressed as pellets. XAS spectra at the Co K-edge were recorded at the SuperXAS beamline of the Swiss Light Source (PSI, Villigen, Switzerland). The incident photon beam provided by a 2.9 T superbend magnet source was collimated by a Si-coated mirror at 2.85 mrad (which also served to reject higher harmonics) and subsequently monochromatized by a Si (111) channel-cut monochromator. A Rh-coated toroidal mirror at 2.5 mrad to focus the X-ray beam to a spot size of 1 mm by 0.2 mm maximal on the sample position was used. The spectra of samples were collected in transmission mode using $N_2$-filled ionization chambers, where a Co foil between the second and third ionization chamber served to calibrate and align all spectra. Commercial CoO and $Co_3O_4$ powders (Aldrich) were used as reference samples.

Density Functional Theory Calculations

Density functional theory (DFT) calculations are used to investigate the electronic properties of the synthesized structures. The calculations are performed using the GPAW code(28,29) and the Atomic Simulation Environment (ASE).(30) Because of the failure of standard DFT in calculating electronic properties in an accurate way, the band gap is estimated by means of non-self-consistent hybrid functional calculations in the framework of the range-separated hybrid functional by Heyd, Scuseria, and Ernzerhof (HSE06).(31,32) The wave functions are expanded in a plane-wave basis with a 500 eV cutoff and a γ-centered Monkhorst–Pack $k$-points grid(33) with a $k$-point density of 3.5 $k$-points Å$^{-1}$. It has been shown that the accuracy of non-self-consistent HSE06 band gaps is rather high and comparable with more computationally expensive many-body GW calculations.(34,35)

Electrochemical Characterization

For the electrochemical characterization, thin-film rotating disk electrode (RDE) measurements were performed.(36,37) Thin-film electrodes were prepared from an ink suspension consisting of 7.5 mg of the oxide powder, 2.5 mL of isopropanol, and 10 μL of Na+-exchanged Nafion as a binder. The ink was sonicated for 30 min and then was dropped and dried on a rotating mirror polished glassy carbon electrode (0.196 cm$^2$). In the present work, the thin-film electrodes are prepared without carbon to obtain an accurate relationship between the oxide properties and the OER activity, since the carbon in a perovskite/carbon composite might play a more complex role than just a simple conductive support.(24−26,38,39) However, for some perovskites, the utilization might strongly be limited by their poor conductivity. Thus, the loading effect on each perovskite powder was studied to find the optimized oxide loading for the OER measurements. (See the Supporting Information, Figure S1.) The OER was investigated using a homemade Teflon cell with a Biologic VMP-300 potentiostat system. The working electrodes were immersed under potential control (1.0 V versus reversible hydrogen electrode, RHE) in a 0.1 M KOH electrolyte at room temperature, and the measurements were performed using a reversible hydrogen electrode (RHE) separated by a salt bridge with diffusion barrier and a gold counter electrode in a three-electrode configuration. The 0.1 M KOH electrolyte was prepared from Milli-Q water and KOH pellets (Sigma-Aldrich, 99.99%). After 30 reverse scan sweeps between 1 and 1.7 V versus RHE at 10 mV s$^{-1}$ and 1600 rpm in synthetic air-saturated electrolyte, chronoamperometry measurements holding each potential for 30 s were performed. The chronoamperometry measurements allow the acquisition of an almost steady-state current with no capacitive contribution (see the Supporting Information, Figure S2). The steady-state current (noted from the chronoamperometry measurement at 1.6 V versus RHE) normalized by the BET surface area of the perovskite oxide was taken as a parameter of OER activity. All the potentials were corrected for the Ohmic-drop in the electrolyte measured by impedance spectroscopy.

For all the samples, the same setup and the same catalyst loading employed for the determination of the OER activity were used to extrapolate the flat-band potential ($E_{fb}$). After immersion of the electrode into the 0.1 M KOH electrolyte, impedance spectroscopy measurements where performed in the frequency range between 10 Hz and 200 kHz, with an ac voltage amplitude of 50 mV, with a DC polarization between 1 and 1.7 V versus RHE. The interfacial capacitance related to the space charge region ($C_{sc}$) has been calculated (i) from the imaginary component of the impedance ($Z''$) at 1 kHz using the relationship $Z'' = 1/2\pi f C_{sc}$, where $f$ is the frequency,(23) and (ii) fitting the obtained Nyquist plots with a $L + R + Q/R$ ($L$ = inductance, $R$ = resistance, $Q$= capacitance) equivalent circuit using Biologic VMP-300 software (see the Supporting Information, Figure S3). The two methods provided very close values. Finally $1/C_{sc}^2$ is traced as a function of applied potential to determine the flat-band potential.

For a measurement of the ex situ conductivity for each material, impedance spectroscopy measurements were performed. The oxide powders were kept under a constant pressure of 0.6 MPa for 5 min, and the electrical resistivity was evaluated by 4-wire impedance spectroscopy measurements at room temperature applying a bias of 100 mV in the frequency range between 1 MHz and 1 Hz. Extrapolating the electrode Ohmic resistance ($R$) from the obtained Nyquist plot, the conductivity ($\sigma$) was calculated as $\sigma = L/RA$ where $L$ and $A$ represent the thickness and the area of the pressed powder, respectively (see the Supporting Information, Figure S4).

## Results and Discussion

### Structure Characterization

Figure 1 shows the X-ray diffraction (XRD) patterns of the investigated perovskites. For most of the perovskites, a pure phase is achieved. As we have discussed in our previous report,(20) there is a phase transition from a rhombohedral ($LaCoO_{3-\delta}$) to a cubic ($La_{0.2}Sr_{0.8}CoO_{3-\delta}$) structure for the $La_{1-x}Sr_xCoO_{3-\delta}$ series as the $x$ value increases. This is due to the Sr substitution which has the effect of straightening the octahedral cage and aligning atoms along the Co–O–Co axis.(20) However, no similar phase transition could be observed for the $LaMO_{3-\delta}$ series as we increase the atomic number of B-site cations from Cr(24) to Ni(28). This indicates that, for the $La_{1-x}Sr_xCoO_{3-\delta}$ series, we could progressively tune the physicochemical properties by only varying the A-site composition. On the other hand, by varying the B-site, the modification of the structural properties among the $LaMO_{3-\delta}$ series is more abrupt and complicated. The PBCO shows a double perovskite structure, and BSCF presents a cubic structure. Figure 1c confirms that the initial carbon in the (BSCF/AB$_f$)$_{centrifuged}$ electrode was removed substantially after the centrifugation process since the XRD pattern of the powder does not show any signal related to an amorphous carbon phase. Figure 1d shows a clear shift of the (BSCF/AB_f)_centrifuged (110) diffraction peak toward lower angles compared to that of BSCF, corresponding to an increase of the lattice parameter from 3.97 to 4.01 Å, respectively. This increase can be ascribed to the reduction of the Co oxidation state in (BSCF/AB_f)_centrifuged compared to that of BSCF, as revealed by our previous XANES reports (see the Supporting Information, Figure S5).(24,25)

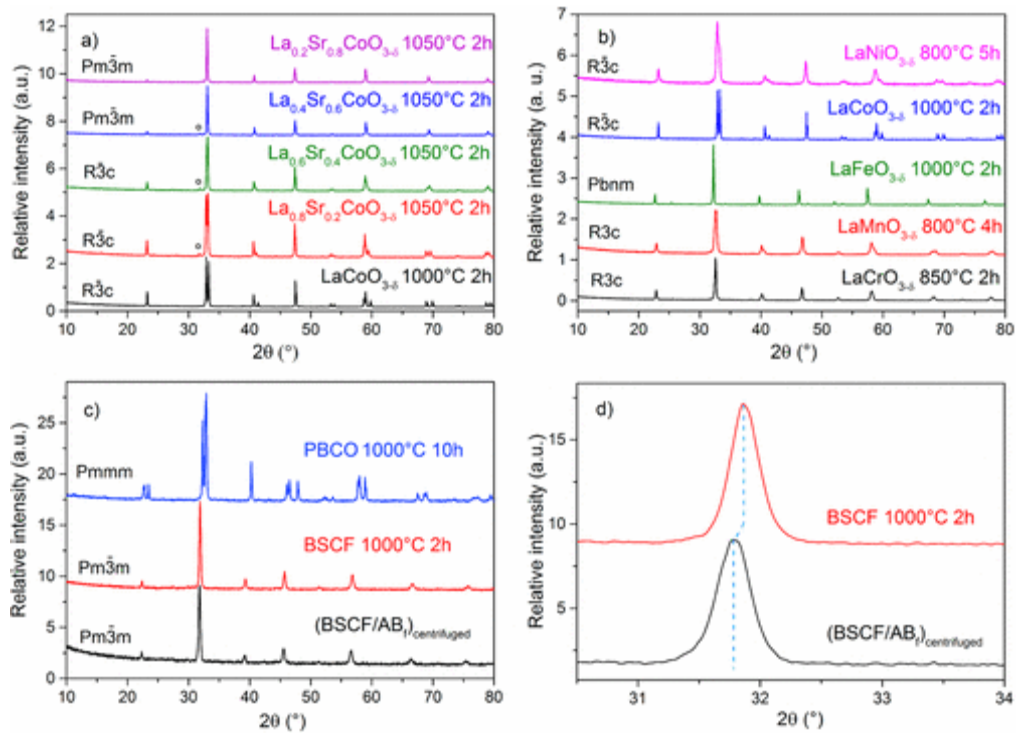

Figure 1. X-ray diffraction (XRD) patterns and group space for the (a) $La_{1-x}Sr_xCoO_{3-\delta}$ series (with $x$ = 0, 0.2, 0.4, 0.6, 0.8), (b) $LaMO_{3-\delta}$ series (M = Cr, Mn, Fe, Co, Ni); (c) BSCF, (BSCF/AB_f)_centrifuged, and PBCO; and (d) (110) diffraction peak for the BSCF and (BSCF/AB_f)_centrifuged powders. The tiny peak denoted by ○ indicates a $(La_ySr_x)_2CoO_4$ secondary phase.

For the acquisition of more precise structural information, specifically the oxygen vacancy content in the perovskites, neutron diffraction measurements have been performed as well. Neutron diffraction is very sensitive to oxygen, and neutron Rietveld refinement allows us to estimate the oxygen vacancy in the perovskite lattice with a precision of 0.01.(40) Figure 2 shows an example of the observed and calculated diffraction patterns for BSCF powder [the neutron diffraction patterns and relative Rietveld refinements of PBCO,

(BSCF/AB$_f$)$_{centrifuged}$, La$_{1-x}$Sr$_x$CoO$_{3-\delta}$, and LaMO$_{3-\delta}$ with M = Cr, Mn, Fe, Co, and Ni are reported in the Supporting Information, Figure S6]. Selected information obtained by the neutron diffraction refinements is presented in Table 1.

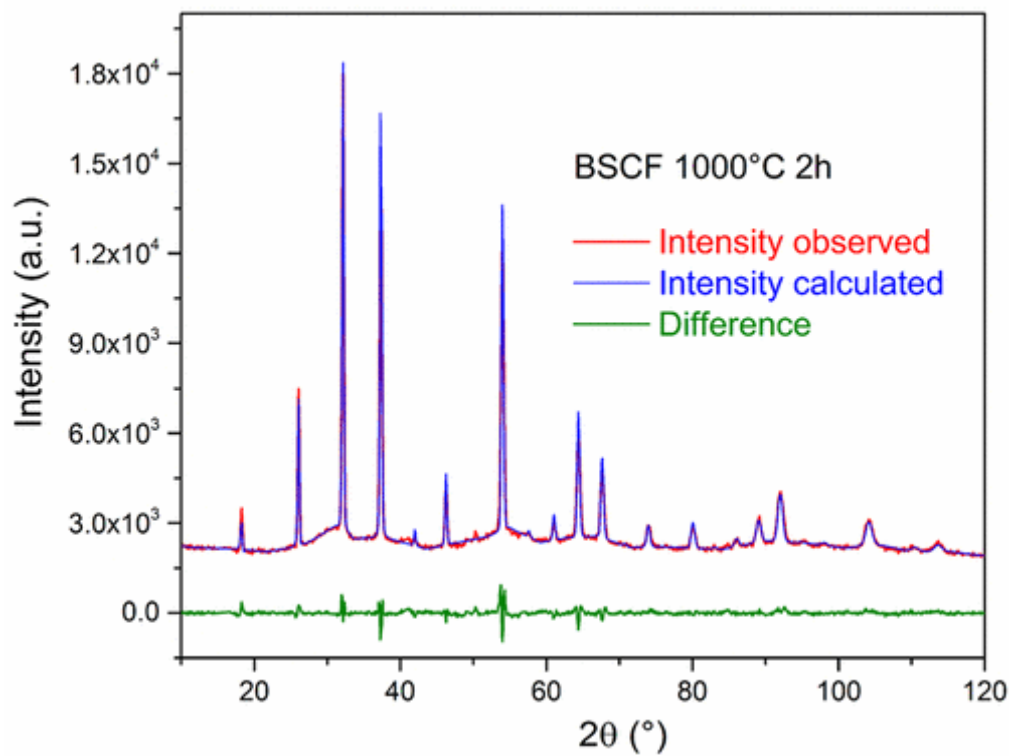

Figure 2. Observed (red line), calculated (blue line), and difference (green line) patterns obtained from the Rietveld refinement of the neutron diffraction data of BSCF. For better fitting of the neutron pattern of BSCF, magnetic peaks have been removed, following the same strategy as in ref (41).

**Table 1. Results of Neutron Diffraction Refinement**

|  | LaCoO$_{3-\delta}$ | La$_{0.8}$Sr$_{0.2}$CoO$_{3-\delta}$ | La$_{0.6}$Sr$_{0.4}$CoO$_{3-\delta}$ | La$_{0.4}$Sr$_{0.6}$CoO$_{3-\delta}$ | La$_{0.2}$Sr$_{0.8}$CoO$_{3-\delta}$ | PBCO[a] |
|---|---|---|---|---|---|---|
| space group | R3c | R3c | R3c | Pm-3m | Pm-3m | Pmmm |
| a (Å) | 5.4118 | 5.4135 | 5.4031 | 3.8080 | 3.8135 | 3.8863 |
| b (Å) |  |  |  |  |  | 7.8020 |
| c (Å) | 13.0175 | 13.0999 | 13.1457 |  |  | 7.6704 |
| M–O–M angle[b] | 161.10° | 165.58° | 168.95° | 180° | 180° | 173.56°/172.04° |
| O vacancy ($\delta$) | 0.01 | 0.01 | 0.06 | 0.05 | 0.14 | 0.24 |

|  | LaCrO$_{3-\delta}$ | LaMnO$_{3-\delta}$ | LaFeO$_{3-\delta}$ | LaNiO$_{3-\delta}$ | BSCF | (BSCF/AB$_f$)$_{centrifuged}$ |
|---|---|---|---|---|---|---|
| space group | R3c | R3c | Pbnm | R3c | Pm-3m | Pm-3m |
| a (Å) | 5.4676 | 5.4660 | 5.5230 | 5.4217 | 3.9730 | 4.0102 |
| b (Å) |  |  | 5.5238 |  |  |  |
| c (Å) | 13.3508 | 13.2609 | 7.8084 | 13.0762 |  |  |
| M–O–M angle[b] | 158.74° | 162.65° | 165.25°/150.45° | 165.69° | 180° | 180° |
| O vacancy ($\delta$) | 0.03 | 0.01 | 0.01 | 0.14 | 0.66 | 0.75 |

a. For a better comparison of the PBCO double perovskite with other single perovskites, the oxygen vacancy content of PBCO is calculated with the Pr$_{0.5}$Ba$_{0.5}$CoO$_{3-\delta}$ formula.

b. M–O–M represents the metal–oxygen–metal angle along chains of MO$_6$ octahedra.

For the La$_{1-x}$Sr$_x$CoO$_{3-\delta}$ series, generally, the oxygen vacancy concentration increases as a function of x value except for La$_{0.4}$Sr$_{0.6}$CoO$_{3-\delta}$. This deviation might be explained by the phase transition from La$_{0.6}$Sr$_{0.4}$CoO$_{3-\delta}$ (rhombohedral) to La$_{0.4}$Sr$_{0.6}$CoO$_{3-\delta}$ (cubic). In fact, when the x value increases from 0.4 to 0.6, the rearrangement of the atoms in the structure might affect the generation of oxygen vacancies. In our previous report,[20] we have demonstrated within our XRD refinement that the Co–O–Co angle along chains of CoO$_6$ octahedra increases as a function of x for the La$_{1-x}$Sr$_x$CoO$_{3-\delta}$ series. In the present work, with the neutron diffraction refinement, the same trend has been obtained. Concerning the LaMO$_{3-\delta}$ series, only LaNiO$_{3-\delta}$ presents relatively high oxygen vacancy content. BSCF and PBCO both present a large amount of oxygen vacancies. Especially for BSCF, after being functionalized by carbon, more oxygen vacancies are generated. This result is consistent with our previous

reports,(24,25) showing a reduced Co oxidation state for the BSCF/carbon composite electrodes (see the Supporting Information, Figure S5).

Ex Situ Electronic Conductivity and Electronic Structure

Figure 3a shows the ex situ electronic conductivity of the La$_{0.8}$Sr$_{0.2}$CoO$_{3-\delta}$ series as a function of the Sr fraction. The ex situ electronic conductivity of LaMO$_{3-\delta}$ powders as a function of the atomic number of the B-site cation is shown in Figure 3b [BSCF, (BSCF/AB$_f$)$_{centrifuged}$, and PBCO are considered as Co-based perovskites]. First, compared to La$_{1-x}$Sr$_x$CoO$_{3-\delta}$ powders where the conductivity increases as a function of the Sr content (Figure 3a), there is no correlation between the conductivity and the atomic number of the B-site cations for the LaMO$_{3-\delta}$ series. Second, for Co-based perovskites (LaCoO$_{3-\delta}$, PBCO, and BSCF), the conductivity seems to be strongly affected by the A- and/or B-site substitution and by the crystal structure. Finally, processing with carbon has a very limited effect on the conductivity for BSCF powders.

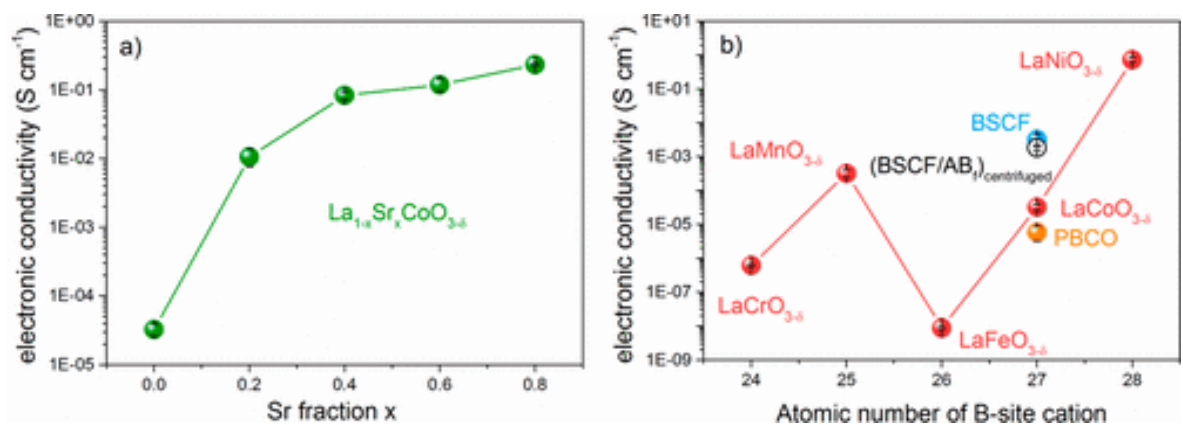

Figure 3. Ex situ electronic conductivity of (a) La$_{1-x}$Sr$_x$CoO$_{3-\delta}$ powders as a function of the Sr fraction and (b) LaMO$_{3-\delta}$ powders as a function of atomic number of B-site cations [BSCF, (BSCF/AB$_f$)$_{centrifuged}$, and PBCO are considered as Co-based perovskites].

In our previous report,(20) we discussed that LaCoO$_{3-\delta}$ oxide is a charge transfer (CT) insulator due to its relatively wide band gap ($\Delta$) between the occupied O 2p valence bands and the unoccupied Co 3d conduction bands. The DFT calculations have shown that the Sr substitution reduces the band gap ($\Delta$) and improves the ex situ electronic conductivity. The same calculations have been performed for the LaMO$_3$ series, BSCF, and PBCO

(see the Supporting Information, Figure S7). The ex situ electronic conductivity as a function of the value of the band gap (Δ) for all the perovskites in this study is shown in Figure 4.

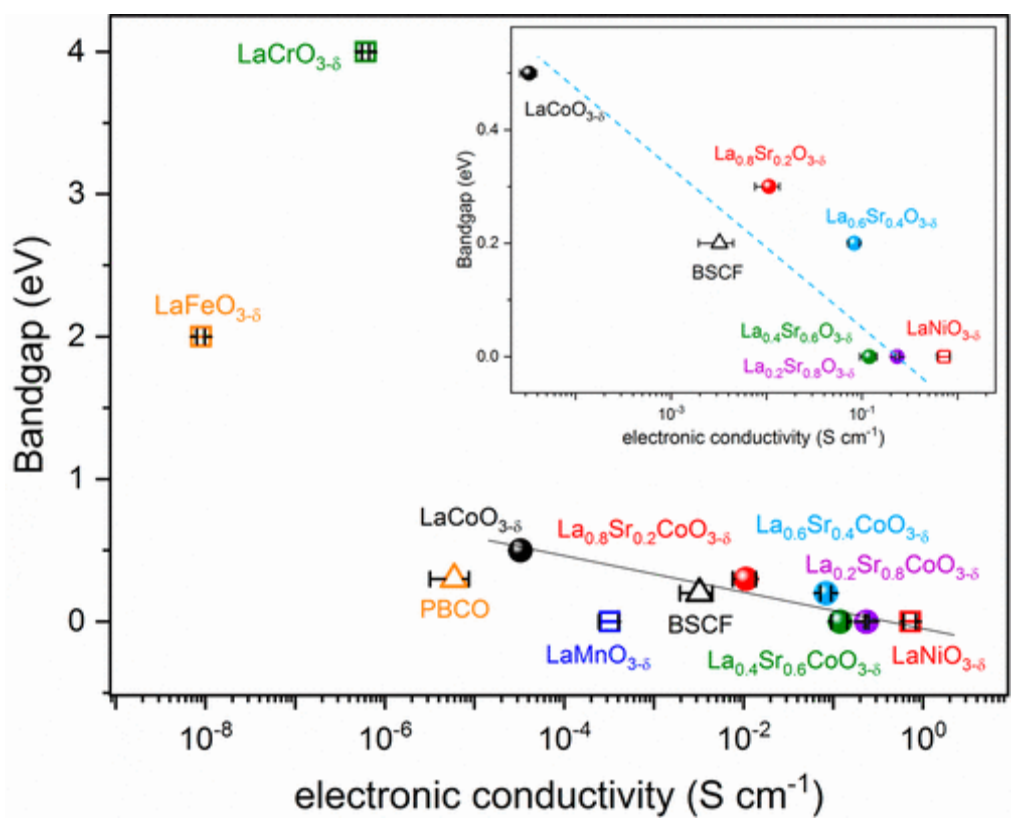

Figure 4. Ex situ electronic conductivity of perovskites measured experimentally as a function of the value of band gap (Δ) deduced from DFT calculations. The inset shows a magnification for the e-linear relationship observed for the $La_{1-x}Sr_xCoO_{3-\delta}$, series, BSCF, and $LaNiO_{3-\delta}$ samples.

Generally, the conductivity increases as the band gap decreases. However, two evident deviation points ($LaFeO_{3-\delta}$ and $LaCrO_{3-\delta}$) and some minor deviations (PBCO, $LaMnO_{3-\delta}$) can be observed compared to the linear relationship between band gap and electronic conductivity observed for the other samples (see inset in Figure 4). These deviations might be due to the limitations related to the DFT calculation. In fact, because of the unit cell size used in the DFT calculations, the exact perovskite stoichiometry and the presence of defects cannot be taken into account.

Previously, it has been shown by DFT calculations(20) that the increase of conductivity as a function of Sr fraction for the $La_{1-x}Sr_xCoO_{3-\delta}$ series is a multieffect consequence of the alignment of the Co–O–Co bonds and the oxidation of the Co cations. However, the effect of the creation of oxygen vacancies on the Co oxidation state was not considered (when $La^{3+}$ is replaced with $Sr^{2+}$, the excess negative charge induced by the Sr substitution could be compensated for by the oxidation of $Co^{3+}$ to higher oxidation state and/or by the creation of oxygen vacancies). In the present work, the oxygen stoichiometry has been determined by neutron diffraction which allows the precise calculation of the oxidation state of the B-site cations. Figure 5a shows the calculated oxidation state of Co cations for the $La_{1-x}Sr_xCoO_{3-\delta}$ series. The Co oxidation state increases as a function of Sr fraction which is consistent with our previous study.(20) For all the $La_{1-x}Sr_xCoO_{3-\delta}$ powders, the calculated value is lower than the theoretical value for ideal stoichiometric compositions ($\delta = 0$). This result indicates that the excess charge induced by Sr-doping to $LaCoO_{3-\delta}$ is compensated both by the oxidation of Co cations and by the creation of oxygen vacancies. However, the oxidation of Co overcomes the creation of oxygen vacancies at $x < 0.6$, while the creation of oxygen vacancies is predominant at $x > 0.6$, consistent with a previous study.(42) Concerning the $LaMO_{3-\delta}$ powders, only the calculated Ni oxidation state value for $LaNiO_{3-\delta}$ is much lower than the theoretical value for ideal stoichiometric compositions (Figure 5b). For PBCO, where the Pr has an oxidation state of 3+,(43) the calculated Co oxidation state value is very close to the theoretical one. Concerning BSCF, because of the high amount of oxygen vacancies, the Co oxidation state is 2.6+, far from the theoretical value of 4+. After sonication with carbon in isopropanol, the Co oxidation state is further reduced to 2.37+, consistent with our XANES measurements (see the Supporting Information, Figure S5).(24,25)

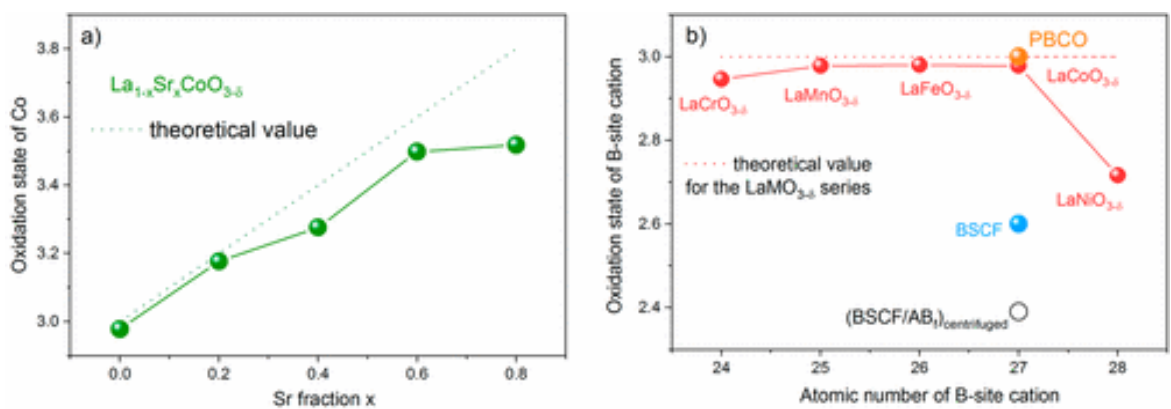

Figure 5. Calculated oxidation state of the (a) Co cation of $La_{1-x}Sr_xCoO_{3-\delta}$ powders as a function of the Sr fraction and (b) B-site cation of $LaMO_{3-\delta}$ powders as a function of atomic number of B-site cations [BSCF, $(BSCF/AB_f)_{centrifuged}$, and PBCO are considered as Co-based perovskites]. The oxygen stoichiometry determined by neutron diffraction is used to perform this calculation. The dashed line shows the theoretical value for ideal $LaMO_3$ stoichiometric compositions (δ = 0). For BSCF and $(BSCF/AB_f)_{centrifuged}$, the theoretical Co oxidation state for ideal stoichiometric composition is 4+. For PBCO the theoretical Co oxidation state for ideal stoichiometric composition is 3.5+.

Flat-Band Potential

The flat-band potential ($E_{fb}$) is a very useful quantity in photoelectrochemistry.(44−48) The first and almost unique study correlating the $E_{fb}$ and the OER activity by J. O. Bockris et al. dates back to 1984.(49) Surprisingly, they could show that there was no correlation between $E_{fb}$ and OER activity for the investigated perovskite oxides. Since then no other studies have been carried out to investigate if a relationship does exist between $E_{fb}$ and OER activity of other perovskite catalysts. Despite this, the $E_{fb}$ is an interesting parameter which can provide fundamental information about the electrochemical interface between the electrode and the electrolyte. Thus, we decided to give the $E_{fb}$ a second chance and try to verify if some correlation exits between the $E_{fb}$ and the OER activity. Briefly, the electrochemical potential of the electrolyte is determined by the redox potential of the solution, and the redox potential of the electrode material is determined by the Fermi level. If the Fermi level and the redox potential of the electrolyte are not equal, there would be a movement of charge between the electrode and the electrolyte to equilibrate the two phases. The excess of charge, which can have a different extension from the interface to the bulk of the electrode (up to 100–10 000 Å for a semiconductor), is usually referred to as the space charge region. Additionally, the bulk Fermi level of the electrode material would be shifted when an external potential is applied to the electrode. Thus, movement of charge between the electrode and the electrolyte varies with the applied potential. There are three different situations to be considered for a p-type semiconductor, which is the case for all the perovskites studied in the present work: (i) When the applied potential is more negative than $E_{fb}$, negative charge is generated on the electrode side with a significant distance (up to 100–10 000 Å) extending into the electrode

(Figure 6a). This process decreases the hole concentration for the p-type semiconductor in this region (depletion layer).(23) (ii) At a certain applied potential, the Fermi level lies at the same energy as the electrolyte redox potential (Figure 6b). There is no net transfer of charge, and this potential is therefore referred to as the flat-band potential ($E_{fb}$). (iii) At applied potentials more positive than $E_{fb}$, electrons are transferred from the electrode to the electrolyte to reach equilibrium conditions, generating more holes at the electrode region. This region is referred to as accumulation layer (Figure 6c).

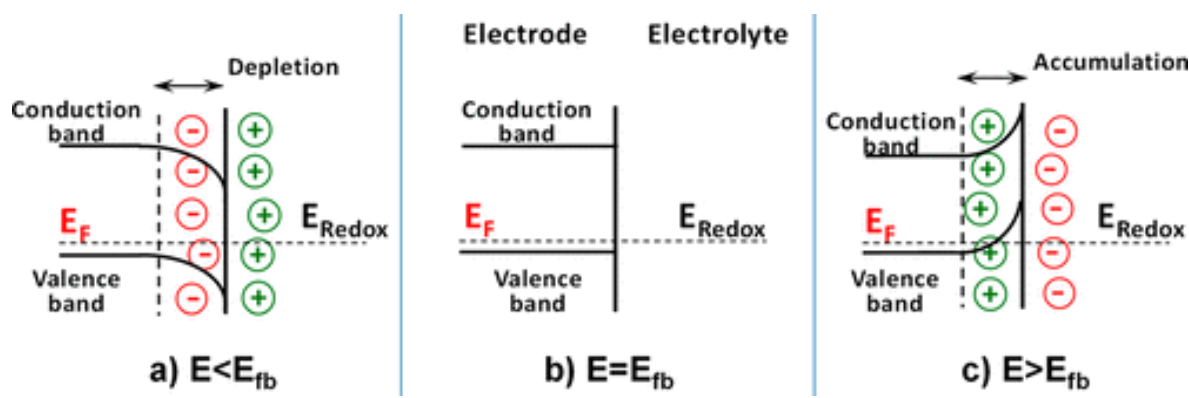

Figure 6. Schematic of a p-type semiconductor showing the effect of varying the applied potential (E). (a) $E < E_{fb}$, (b) $E = E_{fb}$, and (c) $E > E_{fb}$.

Our assumption is that the formation of an accumulation layer at the electrochemical interface would favor the OER. In fact, when the OER occurs, the electrons generated by the reaction need to be transferred from the electrolyte side to the electrode side. Compared to a hole depletion layer, the accumulation layer can favor this transfer, and thus the OER. Furthermore, considering that the minimum OER overpotential calculated by DFT calculations is around 0.3 V,(12) the lowest applied potential for the OER to take place would be around 1.53 V versus RHE. Thus, theoretically, a good p-type semiconductor electrocatalyst should have an $E_{fb}$ lower than 1.53 V versus RHE to ensure the presence of an accumulation layer during the oxygen evolution process.

The flat-band potential of all the perovskites in this study has been measured experimentally with in situ impedance spectroscopy (IS) at different applied potentials. By calculating the electrode capacitance from impedance spectroscopy measurements as a function of applied potentials, it is possible through the Mott–Schottky relation to determine $E_{fb}$ (eq 1):(50−52)

$$\frac{1}{C_{sc}^2} = \frac{2}{\varepsilon\varepsilon_0 N_a}\left(E - E_{fb} - \frac{kT}{e}\right)$$

(1) where $C_{sc}$ is the interfacial capacitance related to the space charge region, ε the dielectric constant of the semiconductor, ε° the permittivity of free space, Na the electron acceptor concentration for a p-type semiconductor, and $E$ the applied potential.

Figure 7 shows an example of a Mott–Schottky plot for $La_{0.8}Sr_{0.2}CoO_{3-\delta}$ powder. The $E_{fb}$ can be determined by extrapolation to $C_{sc}$ = 0. The relationship between the $E_{fb}$ and the OER activity will be discussed in the next section.

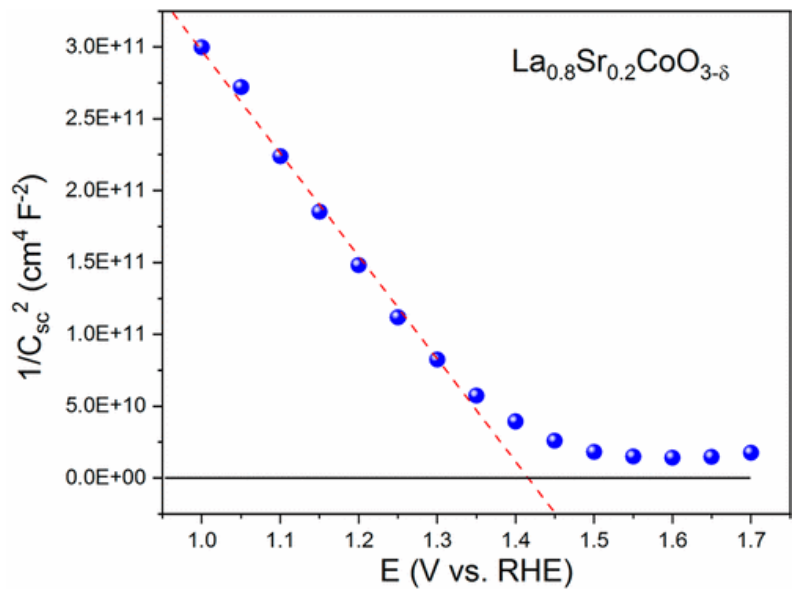

Figure 7. Mott–Schottky plot for $La_{0.8}Sr_{0.2}CoO_{3-\delta}$ powder.

OER–Physicochemical Property Correlation

To discuss the relationship between the OER activity and the physicochemical properties mentioned above, here we trace the current density normalized by the BET surface area of perovskites (μA $cm_{oxide}^{-2}$) as a function of ex situ electronic conductivity (Figure 8a), oxygen vacancy content (Figure 8b), and $E_{fb}$ (Figure 8c).

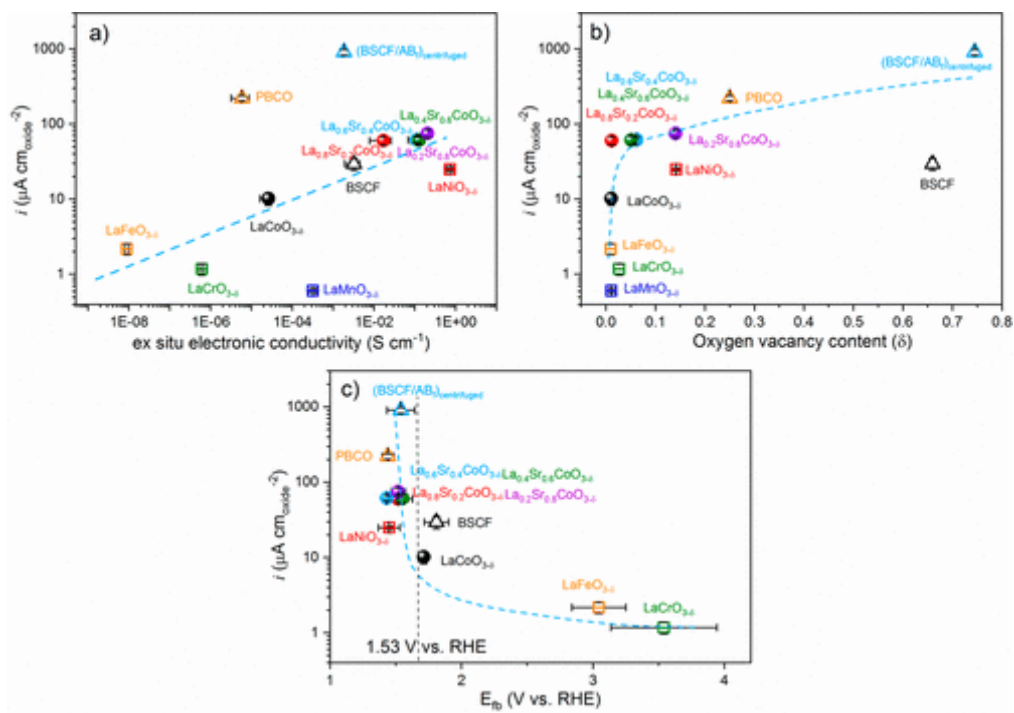

Figure 8. Current density μA cm$_{oxide}^{-2}$ (measured by chronoamperometry measurement at 1.6 V vs RHE) as a function of (a) ex situ electronic conductivity, (b) oxygen vacancy content, and (c) flat-band potential. The flat-band potential of LaMnO$_{3-\delta}$ is not shown in part c because of large error bars in the data (see the Supporting Information, Figure S8).

In Figure 8a, most of the perovskites still follow the same linear trend as previously observed for the La$_{1-x}$Sr$_x$CoO$_{3-\delta}$ series(20) suggesting that high conductivity generally favors high OER activity. However, some strong deviation points from this trend can be observed. PBCO presents very high OER activity despite its relatively low conductivity; LaMnO$_{3-\delta}$ has a very low OER activity, even though it presents a relatively adequate conductivity. After functionalization with carbon, the conductivity of BSCF is not changed, but its OER activity is significantly improved. All these deviations cannot be explained by only considering the OER/conductivity correlation.

Concerning the OER/oxygen vacancy content relationship, generally, higher oxygen vacancy content favors higher OER activity. Considering for example BSCF, after functionalization with carbon, more oxygen vacancies are generated which improves the OER activity. However, it should be taken into account that BSCF without functionalization presents a relatively modest OER activity in spite of its high amount of oxygen vacancy content. Therefore, it appears that also the OER/oxygen vacancy content correlation cannot fully explain the OER activity trend for all samples, particularly for BSCF.

In the previous section, we discussed that, theoretically, a good electrocatalyst should have an $E_{fb}$ lower than 1.53 V versus RHE. This hypothesis is now confirmed by the experimental data shown in Figure 8c, which are also consistent with the report of J.O. Bockris et al.(49) Indeed, a linear correlation between OER activity and $E_{fb}$ cannot be observed, but a correlation between OER activity and $E_{fb}$ does exist since high $E_{fb}$ values represent a strong limitation for the OER activity. When the $E_{fb}$ of a perovskite does not exceed a certain value (about 1.53 V versus RHE), the catalyst has the potential to show fast OER kinetics, but then, other physicochemical properties play a significant role in determining its OER activity. When the $E_{fb}$ of a perovskite is higher than ∼1.53 V versus RHE, the OER activity is strongly limited even though the other physicochemical properties might favor it.

The major conclusion from the data reported in Figure 8 is that it is impossible to only correlate OER activity with a single physicochemical property; i.e., 2D-type correlations as widely presented in the literature have strong limitations and may be valid only for limited perovskite compositions. If the OER activity is correlated to one physicochemical property at a time, there are always some deviation points for this specific correlation. However, these deviations can be explained considering other physicochemical properties and their correlation with the OER activity. For example, (i) PBCO presents relatively low conductivity, but this disadvantage might be compensated by the advantage that PBCO has a high oxygen vacancy content. Additionally, the $E_{fb}$ of PBCO is lower than 1.53 V versus RHE. (ii) The $LaMnO_{3-\delta}$ has a suitable conductivity, but its OER activity might be strongly limited by its low oxygen vacancy content. (iii) The modest OER activity of BSCF might be due to its high $E_{fb}$, even though its suitable conductivity and the high amount of oxygen vacancies favor the OER activity. After functionalization with carbon, the $E_{fb}$ limitation is reduced; thus, the OER activity is significantly improved. These results remind us that the effect of material physicochemical properties on the OER activity is a multidimensional effect. The different descriptors are not isolated from one to another (since a negative physicochemical property can be compensated by another favorable one for the OER). To take these findings into account and to increase the reliability of the OER/descriptor studies, we profile each perovskite to visually associate the OER activity with the three different physicochemical properties discussed above altogether. The profile of each perovskite allows us to deduce the relationship between the OER activity and several physicochemical properties at the same time. Figure 9 shows that a highly active perovskite toward the OER should at least present (i) suitable conductivity (around $10^{-5}$ S cm$^{-1}$), (ii) high amount of oxygen vacancies ($\delta$ higher than 0.2), and (iii) a low $E_{fb}$ (lower than ∼1.53 V versus RHE). The other advantage of the profile plot is that we could continually add new physicochemical properties which are related to the OER activity. Furthermore, it must also take into account that some perovskites can dynamically reconstruct their surface during the OER, as has been demonstrated for BSCF by operando XAS measurements.[3] In situ or operando investigations of the surface state of a wide series of perovskite are an extremely challenging task, out of the scope of the present work, but the changes (increase or decrease) of the OER current during potential cycling for many of the investigated samples (see the Supporting Information, Figure S9) definitively point toward surface catalyst rearrangement and/or catalyst degradation.

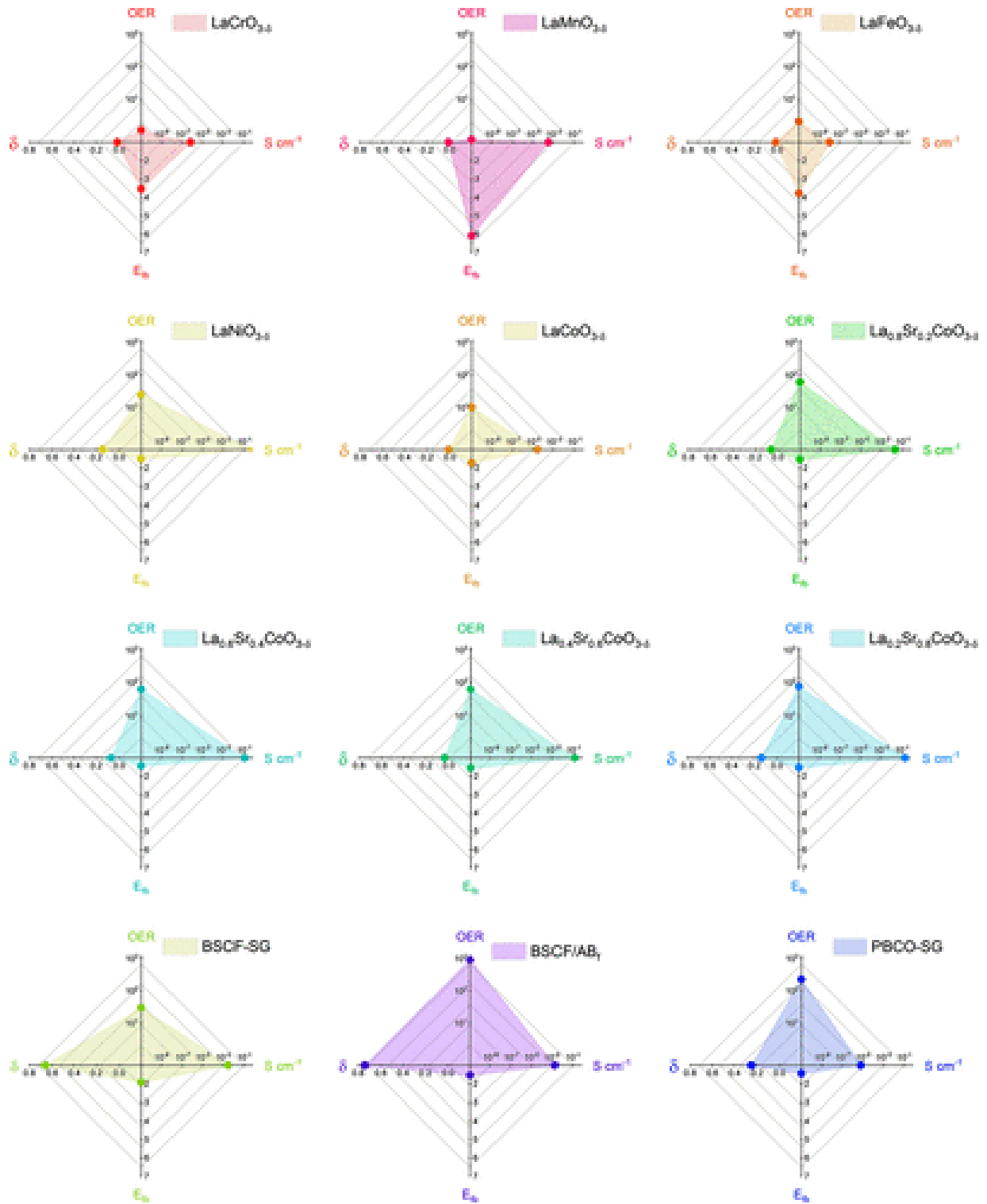

Figure 9. Profile of $La_{1-x}Sr_xCoO_{3-\delta}$, $LaMO_{3-\delta}$, PBCO, BSCF, and $(BSCF/AB_f)_{centrifuged}$ powders combining the OER activity with conductivity (S cm$^{-1}$), amount of oxygen vacancies ($\delta$), and flat-band potential ($E_{fb}$) altogether.

## Conclusion

The correlations between the OER activity and several physicochemical properties (conductivity, oxygen vacancy content, and $E_{fb}$) for a wide range of perovskite compositions [$La_{1-x}Sr_xCoO_{3-\delta}$ series, $LaMO_{3-\delta}$ series, PBCO, BSCF, and BSCF/AB$_f$)$_{centrifuged}$] have been studied experimentally and theoretically. Results show that none of these physicochemical properties alone could be a crucial parameter in determining the electrocatalytic activity of perovskites, pointing to the fact that deviation points are always present if the OER activity of a wide variety of perovskites is correlated to one single physicochemical property at one time. However, the limitation of a 2D correlation (i.e., a correlation between one single property and the OER activity) can be overcome by moving to a combination of descriptors which can better describe a complex scenario as the OER on the surface of oxide materials, where simultaneously to the OER, the metal dissolution and LOER can take place. Therefore, we propose that, for a descriptor study, it is better to consider the relationship of the OER activity with several physicochemical properties at the same time. In fact, it is the correlation of the OER activity with different descriptors that reduces the limitation of a 2D OER/descriptor correlation, leading to a better understanding of the key material properties for highly active OER catalysts. Furthermore, a multidescriptor correlation can also increase the probability to catch false positives, i.e., incidental correlations between a single descriptor and the OER activity for a limited series of catalysts.

According to our study, a highly active perovskite oxide toward the OER should at least present suitable conductivity (around $10^{-5}$ S cm$^{-1}$), high amounts of oxygen vacancies ($\delta$ higher than 0.2), and an $E_{fb}$ lower than ~1.53 V versus RHE. The main finding of the present work opens new perspectives for descriptor-based studies.

## Supporting Information

The Supporting Information is available free of charge on the [ACS Publications website](#) at DOI: [10.1021/acscatal.8b02022](#).

# Acknowledgments

The authors gratefully acknowledge the Swiss National Science Foundation through its Ambizione Program, Innosuisse and the Swiss Competence Center for Energy Research (SCCER) Heat & Electricity Storage, as well as the Swiss National Science Foundation within NCCR Marvel and Paul Scherrer Institute for financial contributions to this work. The authors thank the Swiss Light Source for providing beamtime at the SuperXAS beamline. Authors also thank ILL for provision of neutron beam time (DOI: 10.5291/ILL-DATA.5-22-747) and D1B staff for technical support.

# References

This article references 52 other publications.

**1**

Fabbri, E.; Habereder, A.; Waltar, K.; Kotz, R.; Schmidt, T. J. Developments and Perspectives of Oxide-based Catalysts for the Oxygen Evolution Reaction. *Catal. Sci. Technol.* **2014**, *4*, 3800– 3821, DOI: 10.1039/C4CY00669K

**2**

Herranz, J.; Durst, J.; Fabbri, E.; Patru, A.; Cheng, X.; Permyakova, A. A.; Schmidt, T. J. Interfacial Effects on the Catalysis of the Hydrogen Evolution, Oxygen Evolution and CO2-Reduction Reactions for (co-)electrolyzer Development. *Nano Energy* **2016**, *29*, 4– 28, DOI: 10.1016/j.nanoen.2016.01.027

**3**

Fabbri, E.; Nachtegaal, M.; Binninger, T.; Cheng, X.; Kim, B.; Durst, J.; Bozza, F.; Graule, T.; Schäublin, R.; Wiles, L.; Pertoso, M.; Danilovic, N.; Ayers, K. E.; Schmidt, T. J. Dynamic Surface Self-reconstruction is the Key of Highly Active Perovskite Nano-electrocatalysts for Water Splitting. *Nat. Mater.* **2017**, *16*, 925– 933, DOI: 10.1038/nmat4938

**4**

Binninger, T.; Mohamed, R.; Waltar, K.; Fabbri, E.; Levecque, P.; Kötz, R.; Schmidt, T. J. Thermodynamic Explanation of the Universal Correlation Between Oxygen Evolution Activity and Corrosion of Oxide Catalysts. *Sci. Rep.* **2015**, *5*, 12167, DOI: 10.1038/srep12167

**5**

Mefford, J. T.; Rong, X.; Abakumov, A. M.; Hardin, W. G.; Dai, S.; Kolpak, A. M.; Johnston, K. P.; Stevenson, K. J. Water Electrolysis on La1–xSrxCoO3−δ Perovskite Electrocatalysts. *Nat. Commun.* **2016**, *7*, 11053, DOI: 10.1038/ncomms11053

**6**

Grimaud, A.; Diaz-Morales, O.; Han, B. H.; Hong, W. T.; Lee, Y. L.; Giordano, L.; Stoerzinger, K. A.; Koper, M. T. M.; Shao-Horn, Y. Activating Lattice Oxygen Redox Reactions in Metal Oxides to Catalyse Oxygen Evolution. *Nat. Chem.* **2017**, *9*, 457– 465, DOI: 10.1038/nchem.2695

**7**

Yoo, J. S.; Rong, X.; Liu, Y.; Kolpak, A. M. Role of Lattice Oxygen Participation in Understanding Trends in the Oxygen Evolution Reaction on Perovskites. *ACS Catal.* **2018**, *8*, 4628– 4636, DOI: 10.1021/acscatal.8b00612

**8**

Rong, X.; Parolin, J.; Kolpak, A. M. A Fundamental Relationship between Reaction Mechanism and Stability in Metal Oxide Catalysts for Oxygen Evolution. *ACS Catal.* **2016**, *6*, 1153– 1158, DOI: 10.1021/acscatal.5b02432

**9**

Halck, N. B.; Petrykin, V.; Krtil, P.; Rossmeisl, J. Beyond the Volcano Limitations in Electrocatalysis - Oxygen Evolution Reaction. *Phys. Chem. Chem. Phys.* **2014**, *16*, 13682– 13688, DOI: 10.1039/C4CP00571F

**10**

Reier, T.; Pawolek, Z.; Cherevko, S.; Bruns, M.; Jones, T.; Teschner, D.; Selve, S.; Bergmann, A.; Nong, H. N.; Schlogl, R.; Mayrhofer, K. J. J.; Strasser, P. Molecular Insight in Structure and Activity of Highly Efficient, Low-Ir Ir-Ni Oxide Catalysts for Electrochemical Water Splitting (OER). *J. Am. Chem. Soc.* **2015**, *137*, 13031– 13040, DOI: 10.1021/jacs.5b07788

**11**

Bockris, J. O. M.; Otagawa, T. The Electrocatalysis of Oxygen Evolution on Perovskites. *J. Electrochem. Soc.* **1984**, *131*, 290– 302, DOI: 10.1149/1.2115565

**12**

Man, I. C.; Su, H.-Y.; Calle-Vallejo, F.; Hansen, H. A.; Martínez, J. I.; Inoglu, N. G.; Kitchin, J.; Jaramillo, T. F.; Nørskov, J. K.; Rossmeisl, J. Universality in Oxygen Evolution Electrocatalysis on Oxide Surfaces. *ChemCatChem* **2011**, *3*, 1159– 1165, DOI: 10.1002/cctc.201000397

**13**

Matsumoto, Y.; Yamada, S.; Nishida, T.; Sato, E. Oxygen Evolution on La(1-x)SrxFe(1-y)CoyO3 Series Oxides. *J. Electrochem. Soc.* **1980**, *127*, 2360– 2364, DOI: 10.1149/1.2129415

**14**

Suntivich, J.; May, K. J.; Gasteiger, H. A.; Goodenough, J. B.; Shao-Horn, Y. A Perovskite Oxide Optimized for Oxygen Evolution Catalysis from Molecular Orbital Principles. *Science* **2011**, *334*, 1383– 5, DOI: 10.1126/science.1212858

**15**

Grimaud, A.; Carlton, C. E.; Risch, M.; Hong, W. T.; May, K. J.; Shao-Horn, Y. Oxygen Evolution Activity and Stability of Ba6Mn5O16, Sr4Mn2CoO9, and Sr6Co5O15: The Influence of Transition Metal Coordination. *J. Phys. Chem. C* **2013**, *117*, 25926– 25932, DOI: 10.1021/jp408585z

**16**

Calle-Vallejo, F.; Díaz-Morales, O. A.; Kolb, M. J.; Koper, M. T. M. Why Is Bulk Thermochemistry a Good Descriptor for the Electrocatalytic Activity of Transition Metal Oxides?. *ACS Catal.* **2015**, *5*, 869– 873, DOI: 10.1021/cs5016657

**17**

Calle-Vallejo, F.; Inoglu, N. G.; Su, H. Y.; Martinez, J. I.; Man, I. C.; Koper, M. T. M.; Kitchin, J. R.; Rossmeisl, J. Number of outer electrons as descriptor for adsorption processes on transition metals and their oxides. *Chemical Science* **2013**, *4*, 1245– 1249, DOI: 10.1039/c2sc21601a

**18**

Seo, M. H.; Park, H. W.; Lee, D. U.; Park, M. G.; Chen, Z. Design of Highly Active Perovskite Oxides for Oxygen Evolution Reaction by Combining Experimental and ab Initio Studies. *ACS Catal.* **2015**, *5*, 4337– 4344, DOI: 10.1021/acscatal.5b00114

**19**

Lim, T.; Niemantsverdriet, J. W.; Gracia, J. Layered Antiferromagnetic Ordering in the Most Active Perovskite Catalysts for the Oxygen Evolution Reaction. *ChemCatChem* **2016**, *8*, 2968– 2974, DOI: 10.1002/cctc.201600611

**20**

Cheng, X.; Fabbri, E.; Nachtegaal, M.; Castelli, I. E.; El Kazzi, M.; Haumont, R.; Marzari, N.; Schmidt, T. J. Oxygen Evolution Reaction on La1-xSrxCoO3 Perovskites: A Combined Experimental and Theoretical Study of Their Structural, Electronic, and Electrochemical Properties. *Chem. Mater.* **2015**, *27*, 7662– 7672, DOI: 10.1021/acs.chemmater.5b03138

**21**

Ling, Y. C.; Wang, G. M.; Reddy, J.; Wang, C. C.; Zhang, J. Z.; Li, Y. The Influence of Oxygen Content on the Thermal Activation of Hematite Nanowires. *Angew. Chem., Int. Ed.* **2012**, *51*, 4074– 4079, DOI: 10.1002/anie.201107467

**22**

Zhu, Y. L.; Zhou, W.; Chen, Z. G.; Chen, Y. B.; Su, C.; Tade, M. O.; Shao, Z. P. SrNb0.1Co0.7Fe0.2O3-delta Perovskite as a Next-Generation Electrocatalyst for Oxygen Evolution in Alkaline Solution. *Angew. Chem., Int. Ed.* **2015**, *54*, 3897– 3901, DOI: 10.1002/anie.201408998

**23**

Bott, A. W. Electrochemistry of Semiconductors. *Curr. Sep.* **1998**, *17* (3), 87– 91

**24**

Fabbri, E.; Cheng, X.; Schmidt, T. J. Highly Active Ba0.5Sr0.5Co0.8Fe0.2O3-δ Single Material Electrode Towards the Oxygen Evolution Reaction for Alkaline Water Splitting Applications. *ECS Trans.* **2015**, *69*, 869– 875, DOI: 10.1149/06917.0869ecst

Fabbri, E.; Nachtegaal, M.; Cheng, X.; Schmidt, T. J. Bifunctional Electrocatalytic Activity of Ba0.5Sr0.5Co0.8Fe0.2O3-δ/Carbon Composite Electrodes: Insight into the Local Electronic Structure. *Adv. Energy Mater.* **2015**, *5*, 1402033– 1402038, DOI: 10.1002/aenm.201402033

Mohamed, R.; Cheng, X.; Fabbri, E.; Levecque, P.; Kotz, R.; Conrad, O.; Schmidt, T. J. Electrocatalysis of Perovskites: The Influence of Carbon on the Oxygen Evolution Activity. *J. Electrochem. Soc.* **2015**, *162*, F579– F586, DOI: 10.1149/2.0861506jes

Rodríguez-Carvajal, J. Recent Advances in Magnetic Structure Determination by Neutron Powder Diffraction. *Phys. B* **1993**, *192*, 55– 69, DOI: 10.1016/0921-4526(93)90108-I

Enkovaara, J.; Rostgaard, C.; Mortensen, J. J.; Chen, J.; Dulak, M.; Ferrighi, L.; Gavnholt, J.; Glinsvad, C.; Haikola, V.; Hansen, H. A.; Kristoffersen, H. H.; Kuisma, M.; Larsen, A. H.; Lehtovaara, L.; Ljungberg, M.; Lopez-Acevedo, O.; Moses, P. G.; Ojanen, J.; Olsen, T.; Petzold, V.; Romero, N. A.; Stausholm-Møller, J.; Strange, M.; Tritsaris, G. A.; Vanin, M.; Walter, M.; Hammer, B.; Hakkinen, H.; Madsen, G. K. H.; Nieminen, R. M.; Norskov, J.; Puska, M.; Rantala, T. T.; Schiotz, J.; Thygesen, K. S.; Jacobsen, K. W. Electronic Structure Calculations with GPAW: a Real-space Implementation of the Projector Augmented-wave Method. *J. Phys.: Condens. Matter* **2010**, *22*, 253202, DOI: 10.1088/0953-8984/22/25/253202

Mortensen, J. J.; Hansen, L. B.; Jacobsen, K. W. Real-space Grid Implementation of the Projector Augmented Wave Method. *Phys. Rev. B: Condens. Matter Mater. Phys.* **2005**, *71*, 035109, DOI: 10.1103/PhysRevB.71.035109

**30**

Hjorth Larsen, A.; Mortensen, J. J.; Blomqvist, J.; Castelli, I. E.; Christensen, R.; Dulak, M.; Friis, J.; Groves, M. N.; Hammer, B.; Hargus, C.; Hermes, E. D.; Jennings, P. C.; Jensen, P. B.; Kermode, J.; Kitchin, J. R.; Kolsbjerg, E. L.; Kubal, J.; Kaasbjerg, K.; Lysgaard, S.; Maronsson, J. B.; Maxson, T.; Olsen, T.; Pastewka, L.; Peterson, A.; Rostgaard, C.; Schiotz, J.; Schutt, O.; Strange, M.; Thygesen, K. S.; Vegge, T.; Vilhelmsen, L.; Walter, M.; Zeng, Z. H.; Jacobsen, K. W. The Atomic Simulation Environment-a Python Library for Working with Atoms. *J. Phys.: Condens. Matter* **2017**, *29*, 273002, DOI: 10.1088/1361-648X/aa680e

**31**

Heyd, J.; Scuseria, G. E.; Ernzerhof, M. Hybrid Functionals Based on a Screened Coulomb Potential. *J. Chem. Phys.* **2003**, *118*, 8207– 8215, DOI: 10.1063/1.1564060

**32**

Krukau, A. V.; Vydrov, O. A.; Izmaylov, A. F.; Scuseria, G. E. Influence of the Exchange Screening Parameter on the Performance of Screened Hybrid Functionals. *J. Chem. Phys.* **2006**, *125*, 224106, DOI: 10.1063/1.2404663

**33**

Monkhorst, H. J.; Pack, J. D. Special Points for Brillouin-Zone Integrations. *Phys. Rev. B* **1976**, *13*, 5188– 5192, DOI: 10.1103/PhysRevB.13.5188

**34**

Castelli, I. E.; Huser, F.; Pandey, M.; Li, H.; Thygesen, K. S.; Seger, B.; Jain, A.; Persson, K. A.; Ceder, G.; Jacobsen, K. W. New Light-Harvesting Materials Using Accurate and Efficient Bandgap Calculations. *Adv. Energy Mater.* **2015**, *5*, 1400915, DOI: 10.1002/aenm.201400915

**35**

Paier, J.; Marsman, M.; Hummer, K.; Kresse, G.; Gerber, I. C.; Angyan, J. G. Screened Hybrid Density Functionals Applied to Solids. *J. Chem. Phys.* **2006**, *124*, 154709, DOI: 10.1063/1.2187006

**36**

Schmidt, T. J.; Gasteiger, H. A.; Stab, G. D.; Urban, P. M.; Kolb, D. M.; Behm, R. J. Characterization of High-surface Area Electrocatalysts Using a Rotating Disk Electrode Configuration. *J. Electrochem. Soc.* **1998**, *145*, 2354– 2358, DOI: 10.1149/1.1838642

**37**

Suntivich, J.; Gasteiger, H. A.; Yabuuchi, N.; Shao-Horn, Y. Electrocatalytic Measurement Methodology of Oxide Catalysts Using a Thin-Film Rotating Disk Electrode. *J. Electrochem. Soc.* **2010**, *157*, B1263– B1268, DOI: 10.1149/1.3456630

**38**

Liang, Y. Y.; Li, Y. G.; Wang, H. L.; Zhou, J. G.; Wang, J.; Regier, T.; Dai, H. J. Co3O4 Nanocrystals on Graphene as a Synergistic Catalyst for Oxygen Reduction Reaction. *Nat. Mater.* **2011**, *10*, 780– 786, DOI: 10.1038/nmat3087

**39**

Poux, T.; Napolskiy, F. S.; Dintzer, T.; Keranguevena, G.; Istomin, S. Y.; Tsirlina, G. A.; Antipov, E. V.; Savinova, E. R. Dual Role of Carbon in the Catalytic Layers of Perovskite/carbon Composites for the Electrocatalytic Oxygen Reduction Reaction. *Catal. Today* **2012**, *189*, 83– 92, DOI: 10.1016/j.cattod.2012.04.046

**40**

McIntosh, S.; Vente, J. F.; Haije, W. G.; Blank, D. H. A.; Bouwmeester, H. J. M. Oxygen Stoichiometry and Chemical Expansion of Ba0.5Sr0.5Co0.8Fe0.2O3-δ Measured by in Situ Neutron Diffraction. *Chem. Mater.* **2006**, *18*, 2187– 2193, DOI: 10.1021/cm052763x

**41**

Itoh, T.; Nishida, Y.; Tomita, A.; Fujie, Y.; Kitamura, N.; Idemoto, Y.; Osaka, K.; Hirosawa, I.; Igawa, N. Determination of the Crystal Structure and Charge Density of (Ba0.5Sr0.5)(Co0.8Fe0.2)O2.33 by Rietveld Refinement and Maximum Entropy Method Analysis. *Solid State Commun.* **2009**, *149*, 41– 44, DOI: 10.1016/j.ssc.2008.10.020

**42**

Mineshige, A.; Inaba, M.; Yao, T. S.; Ogumi, Z.; Kikuchi, K.; Kawase, M. Crystal Structure and Metal-insulator Transition of La1-xSrxCoO3. *J. Solid State Chem.* **1996**, *121*, 423– 429, DOI: 10.1006/jssc.1996.0058

**43**

Seymour, I. D.; Tarancon, A.; Chroneos, A.; Parfitt, D.; Kilner, J. A.; Grimes, R. W. Anisotropic Oxygen Diffusion in PrBaCo2O5.5 Double Perovskites. *Solid State Ionics* **2012**, *216*, 41– 43, DOI: 10.1016/j.ssi.2011.09.002

**44**

Butler, M. A. Photoelectrolysis and Physical-Properties of Semiconducting Electrode Wo3. *J. Appl. Phys.* **1977**, *48*, 1914– 1920, DOI: 10.1063/1.323948

**45**

Gao, J. K.; Miao, J. W.; Li, P. Z.; Teng, W. Y.; Yang, L.; Zhao, Y. L.; Liu, B.; Zhang, Q. C. A p-type Ti(IV)-based Metal-organic Framework With Visible-light Photo-response. *Chem. Commun.* **2014**, *50*, 3786– 3788, DOI: 10.1039/C3CC49440C

**46**

Liao, L. B.; Zhang, Q. H.; Su, Z. H.; Zhao, Z. Z.; Wang, Y. N.; Li, Y.; Lu, X. X.; Wei, D. G.; Feng, G. Y.; Yu, Q. K.; Cai, X. J.; Zhao, J. M.; Ren, Z. F.; Fang,

H.; Robles-Hernandez, F.; Baldelli, S.; Bao, J. M. Efficient Solar Water-splitting Using a Nanocrystalline CoO Photocatalyst. *Nat. Nanotechnol.* **2014**, *9*, 69– 73, DOI: 10.1038/nnano.2013.272

**47**

Radecka, M.; Rekas, M.; Trenczek-Zajac, A.; Zakrzewska, K. Importance of the Band Gap Energy and Flat Band Potential for Application of Modified TiO2 Photoanodes in Water Photolysis. *J. Power Sources* **2008**, *181*, 46– 55, DOI: 10.1016/j.jpowsour.2007.10.082

**48**

Rothenberger, G.; Fitzmaurice, D.; Gratzel, M. Optical Electrochemistry 0.3. Spectroscopy of Conduction-Band Electrons in Transparent Metal-Oxide Semiconductor-Films - Optical Determination of the Flat-Band Potential of Colloidal Titanium-Dioxide Films. *J. Phys. Chem.* **1992**, *96*, 5983– 5986, DOI: 10.1021/j100193a062

**49**

Bockris, J. O.; Otagawa, T. The Electrocatalysis of Oxygen Evolution on Perovskites. *J. Electrochem. Soc.* **1984**, *131*, 290– 302, DOI: 10.1149/1.2115565

**50**

Almora, O.; Aranda, C.; Mas-Marza, E.; Garcia-Belmonte, G. On Mott-Schottky analysis interpretation of capacitance measurements in organometal perovskite solar cells. *Appl. Phys. Lett.* **2016**, *109*, 173903, DOI: 10.1063/1.4966127

**51**

Cardon, F.; Gomes, W. P. Determination of Flat-Band Potential of a Semiconductor in Contact with a Metal or an Electrolyte from Mott-Schottky Plot. *J. Phys. D: Appl. Phys.* **1978**, *11*, L63– L67, DOI: 10.1088/0022-3727/11/4/003

**52**

Gelderman, K.; Lee, L.; Donne, S. W. Flat-band Potential of a Semiconductor: Using the Mott-Schottky Equation. *J. Chem. Educ.* **2007**, *84*, 685– 688, DOI: 10.1021/ed084p685